\documentclass[aps,prd,eqsecnum,preprint,tightenlines,nofootinbib,showpacs]{revtex4-1}
\usepackage{graphicx,amsmath,latexsym}

\def\bea{\begin{eqnarray}}
\def\eea{\end{eqnarray}}
\def\be{\begin{equation}}
\def\ee{\end{equation}}
\newcommand{\ub}[1]{\underline{#1}}
\newcommand{\ob}[1]{\overline{#1}}
\newcommand{\Pminus}{{\cal P}^-}
\newcommand{\veck}{\vec{k}_\perp}
\newcommand{\veckp}{\vec{k}_\perp^{\,\prime}}

\begin{document}

\title{An application of the light-front coupled-cluster method \\
to the nonperturbative solution of QED}

\author{Sophia S. Chabysheva}
\author{John R. Hiller}
\affiliation{Department of Physics \\
University of Minnesota-Duluth \\
Duluth, Minnesota 55812}

\date{\today}

\begin{abstract}

The new light-front coupled-cluster (LFCC) method for the
nonperturbative solution of Hamiltonian eigenvalue problems
is described and then illustrated in an application to
quantum electrodynamics. The method eliminates any necessity
for a Fock-space truncation and thereby avoids complications
associated with such a truncation.  An LFCC calculation of
the electron's anomalous magnetic moment is formulated for
a truncation that, for simplicity, excludes positrons but
keeps arbitrarily many photons.

\end{abstract}

%

\maketitle

\section{Introduction}
\label{sec:Introduction}

In order to have a better understanding of hadronic physics
within quantum chromodynamics, it is imperative to have
methods for the nonperturbative solution of the theory.
Lattice gauge theory~\cite{Lattice} has been quite
successful in this regard.  The use of Dyson--Schwinger
equations~\cite{DSE} has also born fruit.  Both, however,
are somewhat indirect, in that they do not provide
Minkowski-space wave functions from which one could
compute observables directly.  An alternative that does
deal directly with such wave functions is the light-front
Hamiltonian method~\cite{DLCQreview}.  There wave functions
appear as coefficients in Fock-state expansions of the
eigenstates.

The light-front Hamiltonian approach now has long
history~\cite{DLCQreview}, beginning with Dirac's
initial formulation of the coordinate choice~\cite{Dirac}.
A key ingredient in practical calculations has been the
truncation of Fock space.  This yields a finite set of
equations for a finite set of wave functions, which are
usually solved numerically.  

Unfortunately, such truncations cause several difficulties,
including uncanceled divergences~\cite{OnePhotonQED},
broken Ward identities~\cite{OSUQED}, and loss of Lorentz
covariance~\cite{BRS,ChiralLimit} and gauge invariance~\cite{ArbGauge}.
These come from the nonperturbative analog of decomposing a
Feynman diagram into time-ordered contributions and throwing
away those contributions that involve more than some fixed
number of intermediate particles.  Some of these difficulties
can be somewhat mitigated by use of a sector-dependent
parameterization~\cite{Wilson,hb,Karmanov}, but they are
not eliminated and can, in fact, lead to
ill-defined wave functions~\cite{SecDep}.

To avoid these difficulties, we have recently proposed
a light-front coupled-cluster (LFCC) method~\cite{LFCClett}.
Fock space is not truncated.  Instead, an eigenstate is
constructed from a valence state $|\phi\rangle$ by the
action of an exponentiated operator $T$.  With inclusion
of a normalizing factor $\sqrt{Z}$, the eigenstate is
written as $\sqrt{Z}e^T|\phi\rangle$.  The problem is
then to find $|\phi\rangle$ and $T$.  To make the
problem finite, $T$ is truncated, but the exponential
of $T$ is not, so that $e^T|\phi\rangle$ includes
contributions from all relevant Fock sectors.
The method relies on light-front coordinates, to make
Fock-state expansions well-defined and to facilitate
separation of internal and external momenta, and on the
mathematical techniques of the many-body coupled-cluster
method~\cite{CCorigin,CCreviews}, hence the name.

The original coupled-cluster method is
used primarily to solve many-body problems where the
number of particles is large and unchanging.
It was first developed by Coester and K\"ummel~\cite{CCorigin}
for applications to the many-body Schr\"odinger equation 
in nuclear physics.  It was extended to many-electron
problems in molecules by \v{C}i\v{z}ek~\cite{Cizek},
which was eventually followed by extensive development
in quantum chemistry~\cite{CCreviews}. 
The basic idea is to
form an eigenstate as $e^T|\phi\rangle$, where
$|\phi\rangle$ is a product of single-particle states,
as in the Hartree--Fock approximation,
and the terms in $T$ annihilate states in $|\phi\rangle$
and create excited states, to build in correlations.
A finite numerical calculation is then done by
truncating $T$ at some (small) number of excitations.
There has also been a series of applications to field theory~\cite{CC-QFT}.
These rely on Fock-state expansions in equal time and
necessarily focus on the structure of the vacuum.
Particle states are built on this vacuum.  There was
some success in analyzing $\phi^4$ theory in $1+1$ dimensions.

Our purpose here is to illustrate the LFCC method in a concrete
calculation in a gauge theory.  We consider the state of
the dressed electron and its anomalous magnetic moment in
an arbitrary covariant gauge.  The theory is regulated
by the inclusion of Pauli--Villars (PV) photons and 
electrons~\cite{OnePhotonQED,TwoPhotonQED}.  We do truncate
Fock space but only to exclude positrons; all possible
single-electron, multi-photon states are retained.  The
exclusion of positrons is not required by the method,
but is done to simplify the illustration.

We begin in Sec.~\ref{sec:method} with an extended description
of the LFCC method~\cite{LFCClett}.  The application to 
QED is developed in the following two sections; Sec.~\ref{sec:qed}
focuses on the eigenvalue problem, and Sec.~\ref{sec:ae}
on the calculation of the anomalous moment.  The details
of light-front QED in an arbitrary covariant gauge~\cite{ArbGauge}
are discussed in Appendix~\ref{sec:arbgauge}, including a
new analysis of the gauge projection.  Details of
the construction of the effective LFCC Hamiltonian are given in
Appendix~\ref{sec:effH}.

\section{A light-front coupled-cluster method}
\label{sec:method}

We use light-front coordinates~\cite{Dirac,DLCQreview},
where time evolves along $x^+\equiv t+z$.  The spatial
coordinates are $\ub{x}=(x^-,\vec{x}_\perp)$, with $x^-\equiv t-z$
and $\vec{x}_\perp=(x,y)$.  The light-front energy is $p^-\equiv E-p_z$
and momentum $\ub{p}=(p^+,\vec{p}_\perp)$, with $p^+\equiv E+p_z$
and $\vec{p}_\perp=(p_x,p_y)$.  The mass-shell condition $p^2=m^2$
becomes $p^-=(m^2+p_\perp^2)/p^+$.  In a system with total momentum
$\ub{P}$, a constituent with momentum $\ub{p}$ is defined to
have a longitudinal momentum fraction $y\equiv p^+/P^+$ and
relative transverse momentum $\veck=\vec p_\perp-y\vec P_\perp$.
A creation operator $a^\dagger(\ub{p})$ for such a particle will
be written as $a^\dagger(y,\veck;\ub{P})$.  Clearly, we have
$a^\dagger(1,0;\ub{P})=a^\dagger(\ub{P})$.

Given a light-front Hamiltonian $\Pminus$, we wish to solve the
fundamental eigenvalue problem
\be
\Pminus|\psi(\ub{P})\rangle=\frac{M^2+P_\perp^2}{P^+}|\psi(\ub{P})\rangle
\ee
for the eigenmass $M$ and the eigenstate $|\psi(\ub{P})\rangle$,
in a basis where the momentum $\ub{{\cal P}}$ is diagonal with
eigenvalue $\ub{P}$.  The normalization is chosen to be
\be
\langle\psi(\ub{P}')|\psi(\ub{P})\rangle=\delta(\ub{P}'-\ub{P}).
\ee
The eigenstate can be generically expanded in a Fock basis
$\{|n;\ub{p_i}\rangle\}$ as
\be
|\psi(\ub{P})\rangle
   =\sum_n\int \left(\prod_i dy_i d\vec{k}_{i\perp}\right)
       \delta(1-\sum_i y_i)  \delta(\sum_i \vec{k}_{i\perp})
       \psi_n(y_i,\vec{k}_{i\perp})|n;\ub{p}_i\rangle,
\ee
where $n$ is the number of constituents and $\ub{p}_i$ their light-front
momenta.  The eigenvalue problem then becomes a coupled set of
integral equations for the wave functions $\psi_n$.  In this form,
the problem requires truncation to a finite set of wave functions,
and the various complications follow, as discussed in the Introduction.

To avoid truncation of the Fock space, we construct the eigenstate as
\be
|\psi(\ub{P})\rangle=\sqrt{Z}e^T|\phi(\ub{P})\rangle.
\ee
The valence state $|\phi(\ub{P})\rangle$ contains a small number
of constituents and will be obtained from diagonalization of an
effective Hamiltonian in the corresponding valence sector, where
the small number of particles is fixed.  The operator $T$ is 
constructed to preserve all the quantum numbers of the valence
state, including momentum $\ub{P}$ and angular momentum 
projection $J_z$, and to always increase the number of constituents.
The exponential of $T$ then generates all the higher Fock states
and their associated wave functions.  The normalizing factor
$\sqrt{Z}$ is chosen to allow the valence state to have the same
momentum normalization as the full state:
\be
\langle\phi(\ub{P}')|\phi(\ub{P})\rangle=\delta(\ub{P}'-\ub{P}).
\ee

The eigenvalue problem can now be written as
\be
\ob{\Pminus}|\phi(\ub{P})\rangle=\frac{M^2+P_\perp^2}{P^+}|\phi(\ub{P})\rangle,
\ee
with $\ob{\Pminus}\equiv e^{-T} \Pminus e^T$ the effective Hamiltonian.
By introducing a projection $P_v$ onto the valence sector,
defined as the sector of Fock space containing the Fock states
found in the valence state, we separate the fundamental
eigenvalue problem into a valence eigenvalue problem
\be  \label{eq:valenceeigenproblem}
P_v\ob{\Pminus}|\phi(\ub{P})\rangle=\frac{M^2+P_\perp^2}{P^+}|\phi(\ub{P})\rangle
\ee
to be solved for the valence state, and a set of auxiliary equations
\be
(1-P_v)\ob{\Pminus}|\phi(\ub{P})\rangle=0
\ee
that determine the operator $T$.  The equations must be solved
simultaneously, of course.

For an exact solution, this system of equations will not be finite;
the $T$ operator will have an infinite number of terms, and there
will be a correspondingly infinite number of auxiliary equations
to fix them.  We therefore truncate $T$ to a few terms and truncate
the projection $1-P_v$ to include only enough higher Fock states to have
enough equations to solve for the terms in $T$.  The construction
of these equations is aided by use of the Baker--Hausdorff expansion
\be \label{eq:BHexpansion}
\ob{\Pminus}=\Pminus+[\Pminus,T]+\frac12[[\Pminus,T],T]+\cdots,
\ee
which can be truncated to the finite number of terms relevant
for the truncated projection.  The restriction that
$T$ always increase particle number is critical for this
truncation of the Baker--Hausdorff expansion.  The exponential
of $T$ is not truncated, so that $e^T|\phi\rangle$ still
contains all the higher Fock states.

The infinity of Fock states included in $e^T|\phi(\ub{P})\rangle$
makes the normalization step nontrivial.  Direct computation
of $Z$ requires an infinite sum.  However, we can still
compute expectation values of observables, using a trick
borrowed from the original coupled-cluster method~\cite{CCreviews}.
Let $\hat O$ be the Hermitian operator representing the
observable of interest.  We wish then to compute
\be
\langle\hat O\rangle=Z\langle\phi(\ub{P})| e^{T^\dagger}\hat O e^T|\phi(\ub{P})\rangle.
\ee
To do this, we define
\be
\ob{O}=e^{-T}\hat O e^T
\ee
and 
\be
|\widetilde\psi(\ub{P})\rangle=Ze^{T^\dagger}e^T|\phi(\ub{P})\rangle
    =\sqrt{Z}e^{T^\dagger}|\psi(\ub{P}\rangle,
\ee
so that
\be  \label{eq:expectationvalue}
\langle\hat{O}\rangle=\langle\widetilde\psi(\ub{P})|\ob{O}|\phi(\ub{P})\rangle.
\ee
By construction, we have
\be \label{eq:tildepsinorm}
\langle\phi(\ub{P}')|\widetilde\psi(\ub{P})\rangle
=\langle\psi(\ub{P}')|\psi(\ub{P})\rangle=\delta(\ub{P}'-\ub{P})
\ee
and
\be
\ob{\Pminus}^\dagger|\widetilde\psi(\ub{P})\rangle
=e^{T^\dagger}\Pminus e^{-T^\dagger}\sqrt{Z}e^{T^\dagger}|\psi(\ub{P})\rangle
=\sqrt{Z}e^{T^\dagger}\Pminus|\psi(\ub{P})\rangle
=\frac{M^2+P_\perp^2}{P^+}|\widetilde\psi(\ub{P})\rangle.
\ee
Thus, we can find $|\widetilde\psi(\ub{P})\rangle$ by solving the
eigenvalue problem for $\ob{\Pminus}^\dagger\neq\ob{\Pminus}$, at 
the same mass $M$, and normalizing $|\widetilde\psi(\ub{P})\rangle$
to the valence state $|\phi(\ub{P})\rangle$ according to
Eq.~(\ref{eq:tildepsinorm}).  When $T$ is truncated, $|\widetilde\psi(\ub{P})\rangle$
must also be truncated~\cite{LFCClett}, to the valence sector
plus those Fock states associated with $T|\phi(\ub{P})\rangle$.
This leaves a finite set of coupled equations for the
functions in $|\widetilde\psi(\ub{P})\rangle$.  Also, the
Baker--Hausdorff expansion of the effective operator $\bar{O}$
can be truncated at a finite number of terms.

One useful generalization of this technique is the calculation
of off-diagonal matrix elements
\be \nonumber
\langle\psi_1(\ub{P}_2)|\hat{O}|\psi_1(\ub{P}_1)\rangle
=\sqrt{Z_1 Z_2}\langle\phi_2(\ub{P}_2)|
     e^{T_2^\dagger}\hat{O}e^{T_1}|\phi_1(\ub{P}_1\rangle.
\ee
This can be done by defining $\ob{O}_i=e^{-T_i}\hat{O}e^{T_i}$
and 
$|\widetilde\psi_i(\ub{P})\rangle
=Z_ie^{T_i^\dagger}e^{T_i}|\phi_i(\ub{P})\rangle$
and then considering the two rearrangements for an Hermitian $\hat{O}$
\be \nonumber
\langle\psi_2(\ub{P}_2)|\hat{O}|\psi_1(\ub{P}_1)\rangle
=\sqrt{\frac{Z_1}{Z_2}}\langle\widetilde\psi_2(\ub{P}_2)|
           \ob{O}_2 e^{-T_2}e^{T_1}|\phi_1(\ub{P}_1)\rangle
\ee
and
\be \nonumber
\langle\psi_1(\ub{P}_1)|\hat{O}|\psi_2(\ub{P}_2)\rangle
=\sqrt{\frac{Z_2}{Z_1}}\langle\widetilde\psi_1(\ub{P}_1)|
    \ob{O}_1 e^{-T_1}e^{T_2}|\phi_2(\ub{P}_2)\rangle.
\ee
The unknown normalization factors cancel in the product of the first
by the conjugate of the second, which yields
\bea  \label{eq:offdiag}
\langle\psi_2(\ub{P}_2)|\hat{O}|\psi_1(\ub{P}_1)\rangle
&=&\sqrt{\langle\widetilde\psi_2(\ub{P}_2)|
             \ob{O}_2 e^{-T_2}e^{T_1}|\phi_1(\ub{P}_1)\rangle} \\
&& 
\times\sqrt{\langle\widetilde\psi_1(\ub{P}_1)|
                \ob{O}_1 e^{-T_1}e^{T_2}|\phi_2(\ub{P}_2)\rangle^*}.
\nonumber
\eea
The correct phase can be obtained by checking one or the other of the
individual matrix elements.  The factors of $e^{-T_1}e^{T_2}$ and
$e^{-T_2}e^{T_1}$ can be evaluated with use of power series expansions
for the exponentials or the Zassenhaus expansion~\cite{Zassenhaus}
\be
e^{T_2-T_1}=e^{T_2}e^{-T_1}e^{\frac12[T_2,T_1]}
   e^{\frac13[T_1,[T_2,T_1]]+\frac16[[T_2,T_1],T_2]}\ldots
\ee
Only a finite number of terms will contribute,
because the $T_i$ only increase particle number and the initial and
final states include only a finite set of Fock sectors.  In the special
case of an expectation value, the formula in (\ref{eq:offdiag}) reduces to
(\ref{eq:expectationvalue}).

Another generalization of the technique for expectation values
is to include a projection $P_s$ of the eigenstate onto a subspace.
We have in mind applications to gauge theories where the projection
is onto a physical subspace of states that satisfy the gauge
condition; however, in this Section we will not need to be specific.
Let $|\psi_s(\ub{P})\rangle=\sqrt{Z_s}P_se^T|\phi(\ub{P})\rangle$
be the projected state, normalized such that
\be
\langle\psi_s(\ub{P}')|\psi_s(\ub{P})\rangle=\delta(\ub{P}'-\ub{P}).
\ee
The unprojected state is related by
\be
|\psi_s(\ub{P})\rangle=\sqrt{\frac{Z_s}{Z}}P_s|\psi(\ub{P})\rangle.
\ee
The expectation value of an operator $\hat O$ is given by
\be
\langle\hat O\rangle_s \equiv \langle\psi_s(\ub{P})|\hat O|\psi_s(\ub{P})\rangle
=Z_s\langle\phi(\ub{P})|e^{T^\dagger}P_s^\dagger\hat O P_s e^T|\phi(\ub{P})\rangle.
\ee
On introduction of the unprojected left-hand eigenstate
$|\widetilde\psi(\ub{P})\rangle$, this becomes
\be
\langle\hat O\rangle_s
=\frac{Z_s}{Z}\langle\widetilde\psi(\ub{P})|
    e^{-T}P_s^\dagger\hat O P_s e^T|\phi(\ub{P})\rangle.
\ee
To obtain the ratio $Z_s/Z$, we assume $Z_s$ and $Z$ to be independent
of the total momentum and use
\be
\delta(\ub{P}'-\ub{P})=\langle\psi_s(\ub{P}')|\psi_s(\ub{P})\rangle
=Z_s\langle\phi(\ub{P}')|e^{T^\dagger}P_s^\dagger P_s e^T|\phi(\ub{P})\rangle
=\frac{Z_s}{Z}\langle\widetilde\psi(\ub{P}')|
    e^{-T}P_s^\dagger P_s e^T|\phi(\ub{P})\rangle.
\ee
Integration over $\ub{P}'$ then yields
\be
\frac{Z}{Z_s}=\int d\ub{P}' \langle\widetilde\psi(\ub{P}')|
                       e^{-T}P_s^\dagger P_s e^T|\phi(\ub{P})\rangle.
\ee
We therefore obtain
\be \label{eq:projected}
\langle\hat O\rangle_s
=\frac{\langle\widetilde\psi(\ub{P})|
    e^{-T}P_s^\dagger\hat O P_s e^T|\phi(\ub{P})\rangle}
 {\int d\ub{P}' \langle\widetilde\psi(\ub{P}')|
                       e^{-T}P_s^\dagger P_s e^T|\phi(\ub{P})\rangle}.
\ee
The remaining matrix elements are computed from the known
$|\phi(\ub{P})\rangle$ and $|\widetilde\psi(\ub{P})\rangle$
by first expanding the exponentials, keeping the total
number of $T$ factors consistent with the truncation 
of $T$, carrying out any contractions of annihilation and 
creation operators, and then applying the projections 
$P_s$ and $P_s^\dagger$.  A Baker--Hausdorff expansion is
not useful here, because of the projections, but the truncated 
expansion of the exponentials is equivalent to a truncated
Baker--Hausdorff expansion.

\section{An application to QED}
\label{sec:qed}

To show how the LFCC method works for a gauge theory, 
we consider the dressed-electron
state in QED, truncated to exclude positrons. The theory is regulated
by one Pauli--Villars (PV) electron, with coupling coefficient $\beta_1=1$,
and two PV photons, with coupling coefficients $\xi_1$ and $\xi_2$.
Additional PV fields are not needed if positrons
are absent~\cite{VacPol}.  The construction of the covariant-gauge
light-front Hamiltonian is discussed in Appendix~\ref{sec:arbgauge},
with additional details given in \cite{ArbGauge}.  The valence state is 
\be
|\phi_a^\pm(\ub{P})\rangle=\sum_i z_{ai}^\pm b_{i\pm}^\dagger(\ub{P})|0\rangle,
\ee
and the LFCC eigenstate is
\be
|\psi_a^\sigma(\ub{P})\rangle=\sqrt{Z}e^T|\phi_a^\sigma(\ub{P})\rangle.
\ee
We truncate the $T$ operator to just photon emission from an electron
\be \label{eq:T}
T=\sum_{ijl\sigma s\lambda}\int dy d\veck
\int \frac{d\ub{p}}{\sqrt{16\pi^3}}\sqrt{ p^+}\,t_{ijl}^{\sigma s\lambda}(y,\veck)
a_{l\lambda}^\dagger(y,\veck;\ub{p}) b_{js}^\dagger(1-y,-\veck;\ub{p}) b_{i\sigma}(\ub{p}).
\ee
Surprisingly, this will introduce as much physics as a two-photon truncation
of the Fock space~\cite{TwoPhotonQED}.  The resulting effective
Hamiltonian $\ob{\Pminus}$ is derived in Appendix~\ref{sec:effH}.

\subsection{Right-hand eigenvalue problem}

The projection (\ref{eq:valenceeigenproblem}) of the eigenvalue
problem onto the valence sector yields,
for the effective Hamiltonian in (\ref{eq:EffH}),
\be \label{eq:z}
m_i^2 z_{ai}^\pm +\sum_j I_{ij} z_{aj}^\pm = M_a^2 z_{ai}^\pm,
\ee
with $a=0,1$ and $M_a$ the $a$th eigenmass.  The self-energy
$I_{ij}$ is defined in (\ref{eq:Iji}).  Clearly, the amplitudes
$z_{ai}^\pm$ are actually independent of spin, and the spin index
$\pm$ can be dropped.  We also have left eigenbras
of $P_v\ob{\Pminus}P_v$,
\be
\langle\widetilde\phi_a^\pm(\ub{P})|=\langle0|\sum_i \tilde{z}_{ai} b_{i\pm}(\ub{P}),
\ee
for which the amplitudes satisfy
\be \label{eq:ztilde}
m_i^2 \tilde{z}_{ai} +\sum_j (-1)^{i+j}I_{ji} \tilde{z}_{aj} 
            = M_a^2 \tilde{z}_{ai}.
\ee
These amplitudes are also independent of spin.

The left and right valence eigenvectors for different eigenvalues are orthogonal.
The normalizations are chosen to satisfy
\be
\langle\widetilde\phi_a^{\sigma'}(\ub{P}')|\phi_b^\sigma(\ub{P})\rangle
   =(-1)^a\delta_{ab}\delta_{\sigma\sigma'}\delta(\ub{P}'-\ub{P})
\ee
Notice that the $a=1$ state has negative norm.
We then have
\be
\sum_i (-1)^i \tilde{z}_{ai} z_{bi}=(-1)^a \delta_{ab}.
\ee
We also have an identity matrix in the valence sector
\be
\sum_a (-1)^a z_{ai} \tilde{z}_{aj} = (-1)^i \delta_{ij}.
\ee

Projection of the eigenvalue problem onto the 
one-electron/one-photon sector gives
\bea \label{eq:projection}
\lefteqn{\sum_i(-1)^i z_{ai}\left\{\rule{0mm}{0.3in}
  h_{ijl}^{\pm s\lambda}(y,\veck)
+\frac12 V_{ijl}^{\pm s\lambda}(y,\veck) \right.} && \\
&&+\left[\frac{m_j^2+k_\perp^2}{1-y}+\frac{\mu_{l\lambda}^2+k_\perp^2}{y}-m_i^2\right]
                 t_{ijl}^{\pm s\lambda}(y,\veck) \nonumber \\
&&\left. +\frac12\sum_{i'} \frac{I_{ji'}}{1-y} t_{ii'l}^{\pm s\lambda}(y,\veck)
-\sum_{j'}(-1)^{i+j'}t_{j'jl}^{\pm s\lambda}(y,\veck)I_{j'i} \right\}=0.  \nonumber
\eea
To partially diagonalize in flavor, we define
\be
C_{abl}^{\sigma s\lambda}(y,\veck)\equiv\sum_{ij}(-1)^{i+j}
z_{ai} \tilde{z}_{bj} t_{ijl}^{\sigma s\lambda}(y,\veck).
\ee
With analogous definitions for the vertex functions $H$ and 
the vertex-loop correction $V$, and with
\be
I_{bb'}\equiv(-1)^{b'}\sum_{ij}(-1)^i \tilde{z}_{bi} z_{b'j} I_{ij},
\ee
we have
\bea  \label{eq:Ceqn}
\left[M_a^2-\frac{M_b^2+k_\perp^2}{1-y}
                  -\frac{\mu_{l\lambda}^2+k_\perp^2}{y}\right]
   C_{abl}^{\sigma s\lambda}(y,\veck)&=&H_{abl}^{\sigma s\lambda}(y,\veck) \\
&&  
   +\frac12\left[V_{abl}^{\sigma s\lambda}(y,\veck)
      -\sum_{b'}\frac{I_{bb'}}{1-y}C_{ab'l}^{\sigma s\lambda}(y,\veck)\right]. \nonumber
\eea
Here the eigenmass $M_b$ has replaced the bare mass $m_j$ in a natural
way, without invocation of a sector-dependent parameterization.  These 
equations are to be solved simultaneously with the valence-sector
equations, (\ref{eq:z}) and (\ref{eq:ztilde}).  To facilitate the
calculation, the self-energy contribution $I_{bb'}$ and 
the vertex correction $V_{abl}^{\sigma s\lambda}$ can
be expressed directly in terms of the wave functions $C_{abl}^{\sigma s\lambda}$
as
\be
I_{bb'}=(-1)^{b'}\sum_{als\lambda}(-1)^{a+l}\epsilon^\lambda
      \int\frac{dyd\veck}{16\pi^3} \widetilde{H}_{bal}^{\pm s\lambda*}(y,\veck)
                C_{b'al}^{\pm s\lambda}(y,\veck)
\ee
and
\bea
V_{abl}^{\sigma s\lambda}(y,\veck)
&=&\sum_{a'b'l'\sigma's'\lambda'}(-1)^{a'+b'+l'}\epsilon^{\lambda'}
\int \frac{dy' d\veckp}{16\pi^3} \frac{\theta(1-y-y')}{\sqrt{(1-y')(1-y)^3}} \\
&&\times 
\widetilde{H}_{bb'l'}^{ss'\lambda'*}(\frac{y'}{1-y},\veckp+\frac{y'}{1-y}\veck)
C_{a'b'l}^{\sigma' s'\lambda}(\frac{y}{1-y'},\veck+\frac{y}{1-y'}\veckp)
C_{aa'l'}^{\sigma\sigma'\lambda'}(y',\veckp),  \nonumber
\eea
where
\be
\widetilde{H}_{abl}^{\sigma s\lambda}(y,\veck)
  =\sum_{ij}(-1)^{i+j}\tilde{z}_{ai} z_{bj} h_{ijl}^{\sigma s\lambda}(y,\veck).
\ee
The original self-energy contribution $I_{ij}$, defined in (\ref{eq:Iji}),
can be obtained as
\be
I_{ij}=(-1)^j\sum_{bb'}(-1)^b z_{bi}\tilde{z}_{b'j}I_{bb'}.
\ee

The wave functions $C_{abl}^{\sigma s\lambda}$ for different $J_z$ index $\sigma$
are then seen to be related in the same pattern as the vertex functions
$H_{abl}^{\sigma s\lambda}$.  Since the spin and polarization dependence
is not affected by the flavor diagonalizations, this pattern can be read
from the structure of the fundamental vertex functions $h_{abl}^{\sigma s\lambda}$
given in (\ref{eq:VertexFunctions}).  We find
\be
C_{abl}^{-+\lambda}=C_{abl}^{+-\lambda*}, \;\;
C_{abl}^{--\lambda}=-C_{abl}^{++\lambda*}, \;\; \mbox{for} \;\; \lambda=\pm,
\ee
\be
C_{abl}^{-+\lambda}=-C_{abl}^{+-\lambda*}, \;\;
C_{abl}^{--\lambda}=C_{abl}^{++\lambda*}, \;\; \mbox{for} \;\; \lambda=0,3.
\ee

\subsection{Left-hand eigenvalue problem}

To obtain the left-hand eigenstates, for use in computation of matrix elements,
we also need to solve 
\be \label{eq:LHeigenproblem}
\ob{\Pminus}^\dagger|\widetilde\psi_a^\sigma(\ub{P})\rangle
    =\frac{M_a^2+P_\perp^2}{P^+}|\widetilde\psi_a^\sigma(\ub{P})\rangle,
\ee
with $M_a$ fixed and $\ob{\Pminus}$ simplified by using $t_{ijl}^{\sigma s\lambda}$
as a solution to Eq.~(\ref{eq:projection}), for which the curly bracket in the
third term of (\ref{eq:EffH}) is zero.  The conjugate is then
\bea \label{eq:EffHconjugate}
\ob{\Pminus}^\dagger&=&\sum_{ijs}(-1)^j\int d\ub{p}
      \left[\delta_{ij}\frac{m_i^2+p_\perp^2}{p^+}+\frac{I_{ij}}{p^+}\right]
          b_{js}^\dagger(\ub{p}) b_{is}(\ub{p}) \\
&& +\sum_{l\lambda}(-1)^l\epsilon^\lambda\int d\ub{p}
          \frac{\mu_{l\lambda}^2+p_\perp^2}{p^+}
             a_{l\lambda}^\dagger(\ub{p}) a_{l\lambda}(\ub{p}) \nonumber  \\
&&+\sum_{ijl\sigma s\lambda}\int dy d\veck 
   \int\frac{d\ub{p}}{\sqrt{16\pi^3p^+}}
         h_{ijl}^{\sigma s\lambda}(y,\veck)
    a_{l\lambda}^\dagger(y,\veck;\ub{p})
   b_{js}^\dagger(1-y,-\veck;\ub{p}) b_{i\sigma}(\ub{p}) \nonumber \\
&& +\sum_{ijl\sigma s \lambda} (-1)^j\sum_{i'l'\sigma'\lambda'}
\int dy d\veck \int\frac{d\ub{p}}{\sqrt{16\pi^3 p^+}}
\int dy' d\veckp \int \frac{d\ub{p}'}{\sqrt{16\pi^3}}\sqrt{p^{\prime +}} \nonumber \\
&& \rule{0.39in}{0mm}\times 
\delta((1-y)p^+-(1-y')p^{\prime +})
\delta((1-y)\vec p_\perp-\veck-(1-y')\vec p_\perp^{\,\prime}+\veckp) \nonumber \\
&& \rule{0.39in}{0mm}\times 
h_{ijl}^{\sigma s\lambda}(y,\veck)t_{i'jl'}^{\sigma' s\lambda'*}(y',\veckp)
a_{l\lambda}^\dagger(y,\veck;\ub{p})
b_{i'\sigma'}^\dagger(\ub{p}') b_{i\sigma}(\ub{p})
a_{l'\lambda'}(y',\veckp;\ub{p}')
    \nonumber \\
&& -\sum_{ijl\sigma s\lambda}(-1)^i \sum_{j'l's'\lambda'}
\int dy d\veck \int \frac{d\ub{p}}{16\pi^3} 
\int dy' d\veckp h_{ijl}^{\sigma s \lambda}(y,\veck)
t_{ij'l'}^{\sigma s'\lambda'*}(y',\veckp) \nonumber \\
&& \rule{0.39in}{0mm}\times 
a_{l\lambda}^\dagger(y,\veck;\ub{p}) 
b_{js}^\dagger(1-y,-\veck;\ub{p}) 
b_{j's'}(1-y',-\veckp;\ub{p})  
a_{l'\lambda'}(y',\veckp;\ub{p}).   \nonumber
\eea
The truncated left-hand eigenvector is
\bea \label{eq:LHket}
|\widetilde\psi_a^\sigma(\ub{P})\rangle&=&|\widetilde\phi_a^\sigma(\ub{P})\rangle  \\
&&+\sum_{jls\lambda}\int dy d\veck\sqrt{\frac{P^+}{16\pi^3}}
l_{ajl}^{\sigma s\lambda}(y,\veck)a_{l\lambda}^\dagger(y,\veck;\ub{P}) 
b_{js}^\dagger(1-y,-\veck;\ub{P})|0\rangle,
\nonumber
\eea
where
$|\widetilde\phi_a^\pm(\ub{P})\rangle=\sum_i \tilde{z}_{ai} b_{i\pm}^\dagger(\ub{P})|0\rangle$
is fixed.

We diagonalize in flavor, defining
\be
D_{abl}^{\sigma s\lambda}(y,\veck)
    \equiv\sum_j(-1)^j z_{bj}^s l_{ajl}^{\sigma s\lambda}(y,\veck),
\ee
\be
J_{ba}^\sigma=(-1)^b\sum_{b'ls\lambda}(-1)^{b'+l}\epsilon^\lambda
\int\frac{dy d\veck}{16\pi^3} C_{bb'l}^{\sigma s\lambda *}(y,\veck)
D_{ab'l}^{\sigma s\lambda}(y,\veck),
\ee
and
\bea
W_{abl}^{\sigma s\lambda}(y,\veck)
&=&\sum_{a'b'l's'\sigma'\lambda'}(-1)^{a'+b'+l'}\epsilon^{\lambda'}
\int \frac{dy' d\veckp}{16\pi^3} \frac{\theta(1-y-y')}{\sqrt{(1-y')^3(1-y)}} \\
&&\times 
C_{bb'l'}^{ss'\lambda'*}(\frac{y'}{1-y},\veckp+\frac{y'}{1-y}\veck)
\widetilde{H}_{a'b'l}^{\sigma' s'\lambda}(\frac{y}{1-y'},\veck+\frac{y}{1-y'}\veckp)
D_{aa'l'}^{\sigma\sigma'\lambda'}(y',\veckp).  \nonumber
\eea
This yields
\bea \label{eq:Deqn}
\left[M_a^2-\frac{M_b^2+k_\perp^2}{1-y}
    -\frac{\mu_{l\lambda}^2+k_\perp^2}{y}\right]
        D_{abl}^{\sigma s\lambda}(y,\veck)&=&\widetilde{H}_{abl}^{\sigma s\lambda}(y,\veck)\\
 &&  +W_{abl}^{\sigma s\lambda}(y,\veck)
     -\sum_{b'} J_{b'a}^\sigma \widetilde{H}_{b'bl}^{\sigma s\lambda}(y,\veck).
     \nonumber
\eea
The right-hand wave functions $C_{abl}^{\sigma s \lambda}$ are input to
this set of equations.  The left-hand wave functions $D_{abl}^{\sigma s \lambda}$
for different $J_z$ index are related in the same way as the right-hand
wave functions:
\be
D_{abl}^{-+\lambda}=D_{abl}^{+-\lambda*}, \;\;
D_{abl}^{--\lambda}=-D_{abl}^{++\lambda*}, \;\; \mbox{for} \;\; \lambda=\pm,
\ee
\be
D_{abl}^{-+\lambda}=-D_{abl}^{+-\lambda*}, \;\;
D_{abl}^{--\lambda}=D_{abl}^{++\lambda*}, \;\; \mbox{for} \;\; \lambda=0,3.
\ee
Both sets of equations, left and right, require numerical techniques
for their solution.

\section{Anomalous magnetic moment}
\label{sec:ae}

We extract the anomalous magnetic moment of the electron
from the spin-flip matrix element of the current 
\be
J^+=\ob{\psi}\gamma^+\psi=2\psi_+\psi_+=2\sum_{ij}\psi_{i+}\psi_{j+}.
\ee
The $\psi_{i+}$ are given by (\ref{eq:psi_i+}).
For our normalization, the general matrix element is~\cite{BrodskyDrell}
\be
16\pi^3\langle\psi_a^\sigma(\ub{P}+\ub{q})|J^+(0)|\psi_a^\pm(\ub{P})\rangle
=2\delta_{\sigma\pm}F_1(q^2)\pm\frac{q^1\pm iq^2}{M_a}\delta_{\sigma\mp}F_2(q^2).
\ee
We compute in the Drell--Yan frame~\cite{DrellYan}, where $q^+=0$ and 
$P^{\prime +}=P^+$, and there are no pair contributions, so that
\be
J^+(0)=2\sum_{ijs}
  \int \frac{d\ub{p}'}{\sqrt{16\pi^3}}\int \frac{d\ub{p}}{\sqrt{16\pi^3}}
    b_{is}^\dagger(\ub{p}') b_{js}(\ub{p}).
\ee

We can apply the formalism for the projected expectation value
(\ref{eq:projected}), with the projection $P_s$ being the projection onto
the physical subspace with the two transverse polarizations $\lambda=1,2$.
The current matrix element is given by
\be 
\langle\psi_a^\sigma(\ub{P}+\ub{q})|J^+(0)|\psi_a^\pm(\ub{P})\rangle
=\frac{\langle\widetilde\psi_a^\sigma(\ub{P}+\ub{q})|
    e^{-T}P_s^\dagger J^+(0) P_s e^T|\phi_a^\pm(\ub{P})\rangle}
 {\int d\ub{P}' \langle\widetilde\psi_a^\pm(\ub{P}')|
                       e^{-T}P_s^\dagger P_s e^T|\phi_a^\pm(\ub{P})\rangle}.
\ee
Note that $T$ is common to both spin projections $\sigma=\pm$ and
flavor eigenstates $a=0$,1; therefore, a general
expression for off-diagonal matrix elements is not needed here.

For the chosen truncation, $e^{\pm T}$ can be replaced by $1\pm T$
and only terms up to first order in $T$ are to be kept.  On 
substitution of (\ref{eq:LHket}) for $|\widetilde\psi_a^\sigma\rangle$,
we have
\bea
\lefteqn{\langle\widetilde\psi_a^\sigma(\ub{P}+\ub{q})|
    e^{-T}P_s^\dagger J^+(0) P_s e^T|\phi_a^\pm(\ub{P})\rangle=\frac{2}{16\pi^3}
    \left[\rule{0mm}{0.3in}
    \delta_{\sigma\pm}(\tilde{z}_{a0}-\tilde{z}_{a1})(z_{a0}-z_{a1})\right.}&& \\
&&  \rule{0.5in}{0mm}  +\sum_{ijj'ls}(-1)^{i+j+j'+l} \int \frac{dy d\veck}{16\pi^3}  \left\{ \sum_{\lambda=\pm}
   l_{ajl}^{\sigma s\lambda *}(y,\veck-y\vec q_\perp)
   t_{ij'l}^{\pm s\lambda}(y,\veck)z_{ai} \right. \nonumber \\
&&  \rule{2.7in}{0mm} 
\left. \left.   -\sum_{\lambda=0}^3 \epsilon^\lambda l_{ajl}^{\sigma s\lambda *}(y,\veck)
   t_{ijl}^{\pm s\lambda}(y,\veck)z_{aj'}\right\}\right] \nonumber
\eea
for the numerator and
\be
\langle\widetilde\psi_a^\pm(\ub{P}')|
                       e^{-T}P_s^\dagger P_s e^T|\phi_a^\pm(\ub{P})\rangle=
   \delta(\ub{P}'-\ub{P})\left[1-\sum_{ijls}(-1)^{i+j+l}\sum_{\lambda=0,3} \epsilon^\lambda
                         l_{ajl}^{\pm s\lambda*}(y,\veck)t_{ijl}^{\pm s\lambda}(y,\veck)
                         z_{ai}\right]
\ee
for the denominator.

We can then extract the form factors (for $a=0$) as 
\bea
F_1(q^2)&=&\frac{1}{\cal N}
  \left[\rule{0mm}{0.3in}(\tilde{z}_{00}-\tilde{z}_{01})(z_{00}-z_{01}) \right.\\
&& +\sum_{ijj'ls}(-1)^{i+j+j'+l}
 \int\frac{dy d\veck}{16\pi^3}
\left\{\sum_{\lambda=\pm} l_{0il}^{\pm s\lambda *}(y,\veck-y\vec{q}_\perp)
        t_{j'jl}^{\pm s\lambda}(y,\veck) z_{0j'}       \right.  \nonumber \\
 && \left. \left. \rule{2in}{0mm}
 -\sum_{\lambda=0}^3\epsilon^\lambda l_{0j'l}^{\pm s \lambda *}(y,\veck) 
             t_{ij'l}^{\pm s \lambda}(y,\veck) z_{0j} \right\}\right]
 \nonumber
\eea
and
\bea
F_2(q^2)&=&\pm\frac{2M_0}{q^1\pm iq^2}\sum_{ijl} (-1)^{i+j+l}
\sum_{j's}(-1)^{j'} \int\frac{dy d\veck}{16\pi^3}  \\
&& \rule{1in}{0mm} \times \frac{1}{\cal N}
    \left\{ \sum_{\lambda=\pm} l_{0il}^{\mp s\lambda *}(y,\veck-y\vec{q}_\perp)
                         t_{j'jl}^{\pm s\lambda}(y,\veck)z_{0j'} \right. \nonumber \\
&& \rule{1.4in}{0mm} \left.
-\sum_{\lambda=0}^3\epsilon^\lambda l_{0j'l}^{\mp s\lambda*}(y,\veck)t_{ij'l}^{\pm s\lambda}(y,\veck)z_{0j}\right\},
            \nonumber
\eea
with
\be
{\cal N}=1-\sum_{ijls}(-1)^{i+j+l}\sum_{\lambda=0,3}\epsilon^\lambda
   \int\frac{dy d\veck}{16\pi^3} l_{0jl}^{\pm s\lambda*}(y,\veck)
          t_{ijl}^{\pm s\lambda}(y,\veck) z_{0i}.
\ee
However, for opposite spins, $l$ and $t$ are orthogonal, because the
azimuthal integration yields
\be
\int dy d\veck l_{aj'l}^{\mp s\lambda*}(y,\veck)t_{ij'l}^{\pm s\lambda}(y,\veck)=0.
\ee
This eliminates the second term in $F_2$.  In the limit that $q^2\rightarrow0$,
the first term can be rewritten as a derivative, as shown in \cite{BrodskyDrell},
to obtain $a_e=F_2(0)$ as
\be
a_e=\pm \frac{M_0}{\cal N}\sum_{ijj'ls}(-1)^{i+j+j'+l}\sum_{\lambda=\pm}
\int \frac{dy d\veck}{16\pi^3} y\, l_{0il}^{\mp s\lambda *}(y,\veck) 
\left(\frac{\partial}{\partial k^1}\mp i\frac{\partial}{\partial k^2}\right)
t_{j'jl}^{\pm s\lambda}(y,\veck) z_{0j'}.
\ee
In terms of the wave functions, $C_{abl}^{\sigma s\lambda}$ and
$D_{abl}^{\sigma s\lambda}$, we have
\be  \label{eq:ae}
a_e=\pm \frac{M_0}{\cal N}\sum_{ijabls}(-1)^{i+j+a+b+l}\tilde{z}_{ai}z_{bj}
   \sum_{\lambda=\pm} \int \frac{dy d\veck}{16\pi^3} 
       y\, D_{0al}^{\mp s\lambda *}(y,\veck) 
\left(\frac{\partial}{\partial k^1}\mp i\frac{\partial}{\partial k^2}\right)
C_{0bl}^{\pm s\lambda}(y,\veck).
\ee
and
\be
{\cal N}=1-\sum_{bls}(-1)^{b+l}\sum_{\lambda=0,3}\epsilon^\lambda
   \int\frac{dy d\veck}{16\pi^3} D_{0bl}^{\pm s\lambda*}(y,\veck)
          C_{0bl}^{\pm s\lambda}(y,\veck).
\ee

In the limit of infinite PV masses, and with $M_0=m_e$ the electron mass,
\bea
F_1(q^2)&=&\frac{1}{\cal N}\left[1+\sum_s\int\frac{dy d\veck}{16\pi^3}
\left\{\sum_{\lambda=\pm} 
  l_{000}^{\pm s\lambda *}(y,\veck-y\vec{q}_\perp)t_{000}^{\pm s\lambda}(y,\veck)
                      \right.\right.\\
 && \left. \left. \rule{1.5in}{0mm}
 -\sum_{\lambda=0}^3\epsilon^\lambda
    l_{000}^{\pm s \lambda *}(y,\veck) t_{000}^{\pm s \lambda}(y,\veck)\right\}\right]
 \nonumber
\eea
and
\be
F_2(q^2)=\pm\frac{2m_e}{q^1\pm iq^2}\frac{1}{\cal N}
\sum_s\sum_{\lambda=\pm}\int\frac{dy d\veck}{16\pi^3} 
    l_{000}^{\mp s\lambda *}(y,\veck-y\vec{q}_\perp)
                         t_{000}^{\pm s\lambda}(y,\veck),
\ee
with
\be
{\cal N}=1-\sum_s\sum_{\lambda=0,3}\epsilon^\lambda
   \int\frac{dy d\veck}{16\pi^3} l_{000}^{\pm s\lambda*}(y,\veck)
          t_{000}^{\pm s\lambda}(y,\veck).
\ee
In the $q^2\rightarrow0$ limit, we have $F_1(0)=1$ and
\be
a_e=\pm \frac{m_e}{\cal N}\sum_s\sum_{\lambda=\pm}
\int \frac{dy d\veck}{16\pi^3}
y\, l_{000}^{\mp s\lambda *}(y,\veck) 
\left(\frac{\partial}{\partial k^1}\mp i\frac{\partial}{\partial k^2}\right)
t_{000}^{\pm s\lambda}(y,\veck).
\ee

The leading perturbative result is
\be
t_{000}^{\sigma s\lambda}(y,\veck)
=l_{000}^{\sigma s\lambda}(y,\veck)
=\frac{h_{000}^{\sigma s\lambda}(y,\veck)}
    {m_e^2-\frac{m_e^2+k_\perp^2}{1-y}
    -\frac{\mu_{0\lambda}^2+k_\perp^2}{y}}.
\ee
When $\mu_0$ goes to zero, we find ${\cal N}=1+{\cal O}(\alpha^2)$ and
\be
a_e=\frac{\alpha}{2\pi}+{\cal O}(\alpha^2),
\ee
which agrees with the Schwinger result~\cite{Schwinger}.
The result for the normalization factor ${\cal N}$ depends
on the fact that $h_{ijl}^{\sigma s\lambda}$ is the
same for $\lambda=0$ and 3 when the photon mass $\mu_l$
is zero, as found in (\ref{eq:VertexFunctions}), so that
the leading ${\cal O}(\alpha)$ contributions to the sum over
$\lambda$ cancel.


\section{Summary}
\label{sec:summary}

We have applied the LFCC method~\cite{LFCClett} to QED in an
arbitrary covariant gauge.  The anomalous moment of the electron
is given by Eq.~(\ref{eq:ae}), which must be evaluated with use
of the left-hand and right-hand functions $D_{ijl}^{\sigma s\lambda}$
and $C_{ijl}^{\sigma s\lambda}$.  These functions are obtained
as solutions of the corresponding eigenvalue problems (\ref{eq:Deqn})
and (\ref{eq:Ceqn}).  The leading perturbative contribution to
the anomalous moment is verified to be the standard Schwinger term;
a nonperturbative solution requires numerical methods.  To carry
out this construction, we have extended our analysis of arbitrary 
gauges~\cite{ArbGauge} to include a specific choice of projection
onto physical states, as discussed at the end of Appendix~\ref{sec:arbgauge},
and have extended the calculation of LFCC matrix elements~\cite{LFCClett}
to include such projections, as expressed in Eq.~(\ref{eq:projected}).

This provides an extensive test of the LFCC method in a gauge theory
regulated by PV fields.  However, the method is much more general than
this; it should be applicable to any regulated light-front Hamiltonian.
The avoidance of a Fock-space truncation then provides several benefits,
with the absence of Fock-sector dependence and spectator dependence
and of the uncanceled divergences that can result from them.  The 
method allows for systematic improvement of a calculation, through
the addition of terms to the exponentiated operator $T$.  Thus a 
natural next step is to include one or more positron terms, to 
study the dressed-photon state~\cite{VacPol}, pair contributions
to the dressed-electron state, and positronium.

\acknowledgments
This work was supported in part by the U.S. Department of Energy
through Contract No.\ DE-FG02-98ER41087.

\appendix

\section{Light-front QED in an arbitrary gauge} \label{sec:arbgauge}

Here we summarize a formulation of light-front QED regulated by the
inclusion of Pauli--Villars (PV) electrons and photons~\cite{PauliVillars}
and quantized in an arbitrary covariant gauge.  Details can be found 
in \cite{ArbGauge}.  We write the Lagrangian as
\bea  \label{eq:Lagrangian}
{\cal L} &=&  \sum_{i=0}^2 (-1)^i \left[-\frac14 F_i^{\mu \nu} F_{i,\mu \nu} 
         +\frac12 \mu_i^2 A_i^\mu A_{i\mu} 
         -\frac12 \zeta \left(\partial^\mu A_{i\mu}\right)^2\right] \\
&& + \sum_{i=0}^2 (-1)^i \bar{\psi_i} (i \gamma^\mu \partial_\mu - m_i) \psi_i 
  - e \bar{\psi}\gamma^\mu \psi A_\mu , \nonumber
\eea
with
\be \label{eq:NullFields}
  \psi =  \sum_{i=0}^2 \sqrt{\beta_i}\psi_i, \;\;
  A_\mu  = \sum_{i=0}^2 \sqrt{\xi_i}A_{i\mu}, \;\;
  F_{i\mu \nu} = \partial_\mu A_{i\nu}-\partial_\nu A_{i\mu} .
\ee
The index $i$ is zero for physical fields and 1 or 2 for PV fields.  The
coupling coefficients $\beta_i$ and $\xi_i$ are constrained to satisfy
$\beta_0=1$, $\xi_0=1$, and
\be
\sum_{i=0}^2(-1)^i\xi_i=0, \;\;
\sum_{i=0}^2(-1)^i\beta_i=0, 
\ee
and to reproduce the correct chiral symmetry in the limit of a massless
electron~\cite{ChiralLimit}, and to guarantee a massless eigenstate for
the photon~\cite{VacPol}.

The fermion fields $\psi_i$ are decomposed into dynamical and
nondynamical parts $\psi_{i\pm}\equiv\Lambda_\pm\psi_i$ by
the complementary projections 
$\Lambda_\pm\equiv\gamma^0\gamma^\pm/2$~\cite{DLCQreview,LepageBrodsky}.
The dynamical part is written
\be \label{eq:psi_i+}
\psi_{i+}=\frac{1}{\sqrt{16\pi^3}}\sum_s\int d\ub{k} \chi_s
  \left[b_{is}(\ub{k})e^{-i\ub{k}\cdot\ub{x}}
        +d_{i,-s}^\dagger(\ub{k})e^{i\ub{k}\cdot\ub{x}}\right],
\ee
with
\be
\chi_+=\frac{1}{\sqrt{2}}\left(\begin{array}{c} 1 \\ 0 \\ 1 \\ 0 \end{array}\right),\;\;
\chi_-=\frac{1}{\sqrt{2}}\left(\begin{array}{c} 0 \\ 1 \\ 0 \\ -1 \end{array}\right),
\ee
and
\bea
\{b_{is}(\ub{k}),b_{i's'}^\dagger(\ub{k}'\}
   &=&(-1)^i\delta_{ii'}\delta_{ss'}\delta(\ub{k}-\ub{k}'), \\
\{d_{is}(\ub{k}),d_{i's'}^\dagger(\ub{k}'\}
   &=&(-1)^i\delta_{ii'}\delta_{ss'}\delta(\ub{k}-\ub{k}').
\eea
The nondynamical part satisfies the constraint
\bea \label{eq:FermionConstraint}
i(-1)^i\partial_-\psi_{i-}&+&e A_-\sqrt{\beta_i}\sum_j\psi_{j-}  \\
  &=&(i\gamma^0\gamma^\perp)
     \left[(-1)^i\partial_\perp \psi_{i+}-ie A_\perp\sqrt{\beta_i}\sum_j\psi_{j+}\right] 
      -(-1)^i m_i \gamma^0\psi_{i+} . \nonumber
\eea
However, the constraint for the sum that enters the interaction becomes
simply
\be
i\partial_-\psi_-
  =(i\gamma^0\gamma^\perp)
     \partial_\perp \psi_+
      - \gamma^0\sum_i\sqrt{\beta_i}m_i\psi_{i+},
\ee
where the photon field does not appear.  The constraint can
then be trivially solved and the nondynamical field removed from
the Lagrangian.

A vector field of mass $\mu_l$ is written as
\bea
A_{l\mu}(x)&=&\int\frac{d\ub{k}}{\sqrt{16\pi^3 k^+}}\left\{\sum_{\lambda=1}^3
   e_\mu^{(\lambda)}(\ub{k})\left[ a_{l\lambda}(\ub{k})e^{-ik\cdot x}
            + a_{l\lambda}^\dagger(\ub{k})e^{ik\cdot x}\right]\right. \\
&& \rule{1in}{0mm}
\left.+e_\mu^{(0)}(\ub{k})\left[ a_{l0}(\ub{k})e^{-i\tilde k\cdot x}
            + a_{l0}^\dagger(\ub{k})e^{i\tilde k\cdot x}\right]\right\}, \nonumber
\eea
with $\tilde k$ a four-vector associated with a different mass
$\tilde\mu_l\equiv\mu_l/\sqrt{\zeta}$, such that
\be
\ub{\tilde k}=\ub{k}, \;\; \tilde k^-=(k_\perp^2+\tilde\mu_l^2)/k^+ .
\ee
This allows the field to satisfy the Euler--Lagrange equation, while 
the fourth polarization $\lambda=0$ does not satisfy the gauge 
condition $\partial\cdot A_l=0$~\cite{ArbGauge}.
The polarization vectors are defined by
\bea
e^{(1,2)}(\ub{k})&=&(0,2 \hat e_{1,2}\cdot \vec{k}_\perp/k^+,\hat e_{1,2}), \\
e^{(3)}(\ub{k})&=&((k_\perp^2-\mu_l^2)/k^+,k^+,\vec k_\perp)/\mu_l, \label{eq:e3} \\
e^{(0)}(\ub{k})&=&\tilde k/\mu_l=((k_\perp^2+\tilde\mu_l^2)/k^+,k^+,\vec k_\perp)/\mu_l,
\label{eq:e0}
\eea
and the commutation relations are
\be  \label{eq:commutationrelations}
[a_{l\lambda}(\ub{k}),a_{l'\lambda'}^\dagger(\ub{k}')]
     =(-1)^l\delta_{ll'}\epsilon^\lambda \delta_{\lambda\lambda'}\delta(\ub{k}-\ub{k}'),
\ee
with $\epsilon=(-1,1,1,1)$ the metric signature for the physical photon.

For the purpose of making $J_z$-conservation explicit, it is convenient
to introduce circular polarizations
\be
e^{(\pm)}=\mp\frac{1}{\sqrt{2}}(e^{(1)}\pm ie^{(2)})
\ee
and the associated creation and annihilation operators
\be
a_{l\pm}^\dagger=\mp\frac{1}{\sqrt{2}}(a_{l1}^\dagger\pm ia_{l2}^\dagger),\;\;
a_{l\pm}=\mp\frac{1}{\sqrt{2}}(a_{l1}\mp ia_{l2}).
\ee
These operators satisfy the same commutation relations (\ref{eq:commutationrelations}),
with $\lambda$ now taking the values $\pm$ instead of 1 and 2, and with
$\epsilon^\pm\equiv1$.  Sums over $\lambda$ will, in general, include
0, $\pm$, and 3.

The light-front Hamiltonian is then given by
\be \label{eq:Pminus}
\Pminus=\Pminus_{0a}+\Pminus_{0b}+\Pminus_{\rm int},
\ee
with
\bea
\Pminus_{0a} &=&   \sum_{l\lambda}(-1)^l\epsilon^\lambda\int d\ub{p}
          \frac{\mu_{l\lambda}^2+p_\perp^2}{p^+}
             a_{l\lambda}^\dagger(\ub{p}) a_{l\lambda}(\ub{p}), \\
\Pminus_{0b} &=&   \sum_{is}(-1)^i\int d\ub{p}
      \frac{m_i^2+p_\perp^2}{p^+}b_{is}^\dagger(\ub{p}) b_{is}(\ub{p}) ,
\eea
and, if antifermion terms are neglected,
\bea
\Pminus_{\rm int} &=&   \sum_{ijl\sigma s\lambda}\int dy d\veck 
   \int\frac{d\ub{p}}{\sqrt{16\pi^3p^+}}   \\
&& \times    \left\{h_{ijl}^{\sigma s\lambda}(y,\veck)
        a_{l\lambda}^\dagger(y,\veck;\ub{p})
           b_{js}^\dagger(1-y,-\veck;\ub{p})b_{i\sigma}(\ub{p}) \right. \nonumber \\
&& \left. +h_{ijl}^{\sigma s\lambda *}(y,\veck)b_{i\sigma}^\dagger(\ub{p})
     b_{js}(1-y,-\veck;\ub{p})a_{l\lambda}(y,\veck;\ub{p}) \right\}.  \nonumber
\eea
As discussed at the beginning of Sec.~\ref{sec:method},
the multiple arguments of the creation and annihilation operators
are defined by $a_{l\lambda}(y,\veck;\ub{p})\equiv 
a_{l\lambda}(yp^+,y\vec{p}_\perp+\veck)$.  The 
vertex functions are
\bea \label{eq:VertexFunctions}
h_{ijl}^{\pm\pm\pm}(y,\veck)&=&
\mp e\sqrt{2\beta_i\beta_j\xi_l}\,\frac{k_\perp e^{\mp i\phi}}{(1-y)y^{3/2}}, \\
h_{ijl}^{\pm\pm\mp}(y,\veck)
&=&\pm e\sqrt{2\beta_i\beta_j\xi_l}\,\frac{k_\perp e^{\pm i\phi}}{y^{3/2}}, \nonumber \\
h_{ijl}^{\mp\pm\pm}(y,\veck)&=&0, \nonumber \\
h_{ijl}^{\mp\pm\mp}(y,\veck)
&=&e\sqrt{2\beta_i\beta_j\xi_l}\,\frac{m_i(1-y)-m_j}{(1-y)\sqrt{y}}, \nonumber \\
h_{ijl}^{\pm\pm 3}(y,\veck)
&=&e\sqrt{\beta_i\beta_j\xi_l}\,\frac{m_im_j y^2-\mu_l^2(1-y)+k_\perp^2}{\mu_l (1-y) y^{3/2}}, 
\nonumber \\
h_{ijl}^{\pm\pm 0}(y,\veck)
&=&e\sqrt{\beta_i\beta_j\xi_l}\,\frac{m_im_j y^2+\tilde\mu_l^2(1-y)+k_\perp^2}
                                   {\mu_l (1-y) y^{3/2}},   \nonumber \\
h_{ijl}^{\mp\pm 3}(y,\veck)
&=&h_{ijl}^{\mp\pm 0}(y,\veck)
=\pm e\sqrt{\beta_i\beta_j\xi_l}\,\frac{(m_j-m_i)k_\perp e^{\mp i\phi}}{\mu_l (1-y)\sqrt{y}}.
\nonumber 
\eea

The gauge condition $\partial\cdot A_l=0$ for the $l$th flavor is implemented
as a projection onto a physical subspace.  Only the polarization with
opposite metric, $\lambda=0$, contributes to $\partial\cdot A_l$; we have
\be
\partial\cdot A_l=-i\frac{\mu_l}{\sqrt\zeta}\int \frac{d\ub{k}}{\sqrt{16\pi^3k^+}}
\left[ a_{l0}(\ub{k})e^{-i\tilde{k}\cdot x}-a_{l0}^\dagger(\ub{k})e^{i\tilde{k}\cdot x}\right].
\ee
We require that physical states $|\psi_{\rm phys}\rangle$ satisfy
\be
\langle\psi_{\rm phys}|\partial\cdot A_l|\psi_{\rm phys}\rangle=0,
\ee
which is guaranteed if $|\psi_{\rm phys}\rangle$ is annihilated by the
positive-frequency part of $\partial\cdot A_l$ or, equivalently,
\be \label{eq:gaugeproj}
\mu_l a_{l0}(\ub{k})|\psi_{\rm phys}\rangle=0.
\ee
However, in the massless limit, the polarization vectors $e^{(3)}$ and
$e^{(0)}$ become identical.  This suggests that implementation of
the gauge projection for a massless photon should involve the removal
of not just the $e^{(0)}$ polarization but also the $e^{(3)}$ 
polarization, as one would expect on physical grounds.

To make this more precise, define two new polarizations
\be
\tilde{e}^{(0,3)}(\ub{k})=\frac{\mu_l}{2k^+}(e^{(3)}+e^{(0)})
                            \pm\frac{k^+}{2\mu_l}(e^{(0)}-e^{(3)})
\ee
before taking the massless limit.  With use of the definitions
in Eqs.~(\ref{eq:e3}) and (\ref{eq:e0}), they can be written
explicitly as
\be
\tilde{e}^{(0,3)}(\ub{k})=\frac{1}{k^+}
   \left(\frac{k_\perp^2+\mu_l^2(1-\zeta)/2\zeta\pm (k^+)^2(1+\zeta)/2\zeta}{k^+},\ub{k}\right).
\ee
These have the useful properties of not being the same in the massless
limit and of reducing, in Feynman gauge ($\zeta=1$), to the standard
choice of~\cite{ChiralLimit}
\be
\tilde{e}^{(3)}=\frac{k-(n\cdot k)n}{n\cdot k}\;\; 
\mbox{and} \;\; \tilde{e}^{(0)}=n,
\ee
where $n$ is the timelike four-vector that reduces to $n=(1,0,0,0)$ in
the frame where $\veck=0$.  By requiring that
\be
\sum_{\lambda=0,3}\tilde{e}^{(\lambda)}(\ub{k})\tilde{a}_{l\lambda}(\ub{k})
=\sum_{\lambda=0,3}e^{(\lambda)}(\ub{k})a_{l\lambda}(\ub{k}),
\ee
we find that
\be
a_{l0}(\ub{k})=\left(\frac{\mu_l}{2k^+}-\frac{k^+}{2\mu_l}\right)\tilde{a}_{l3}
          +\left(\frac{\mu_l}{2k^+}+\frac{k^+}{2\mu_l}\right)\tilde{a}_{l0}
\ee
and that the $\tilde{a}_{l\lambda}$ satisfy the same commutation relations
as do the $a_{l\lambda}$.  

The gauge projection (\ref{eq:gaugeproj}) now becomes, in the massless limit,
\be
(\tilde{a}_{l0}(\ub{k})-\tilde{a}_{l3}(\ub{k}))|\psi_{\rm phys}\rangle=0.
\ee
This removes from $|\psi_{\rm phys}\rangle$ one null combination of 
$\tilde{a}_{l0}$ and $\tilde{a}_{l3}$.  The other null combination,
$\tilde{a}_{l0}+\tilde{a}_{l3}$ remains but makes no
contribution to physical quantities simply because it is null;
instead, it represents the remaining gauge freedom not fixed by
the Lorentz gauge condition~\cite{ChiralLimit}.  For practical calculations
of observables, it too can be removed, which makes the gauge projection
equivalent to the removal of both the $e^{(0)}$ and $e^{(3)}$
polarizations.

For the massive PV photons, we apply the same projection.  This could
not be done for a physical massive vector particle, for which three
polarizations must be retained.  For the PV photons, which are
regulators and not physical, the projection becomes part of the
regularization prescription.  In any case, in the infinite-mass
limit, all such contributions from the PV photons disappear; 
removing them prior to taking this limit will not change the result.

\section{Derivation of the Effective Hamiltonian}
\label{sec:effH}

\begin{figure}[ht]
\vspace{0.2in}
\begin{tabular}{c}
\centerline{\includegraphics[width=12cm]{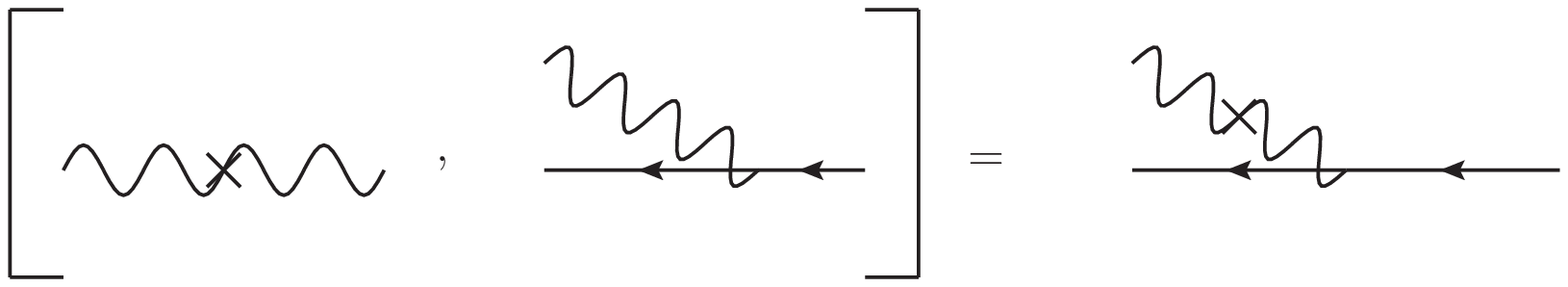}} \\
(a) \\
\centerline{\includegraphics[width=15cm]{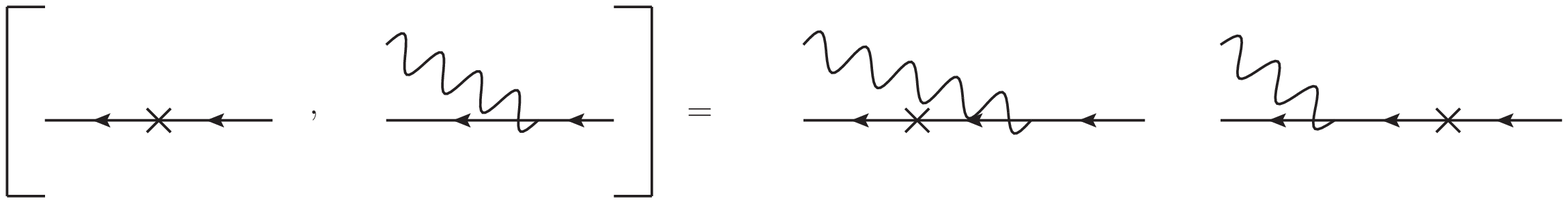}} \\
(b)
\end{tabular}
\caption{\label{fig:P0T} Graphical representations of the
terms in (a) $[\Pminus_{0a},T]$ and (b) $[\Pminus_{0b},T]$.
Each diagram represents an operator that annihilates particles
on the right and creates particles on the left.  The crosses
represent kinetic-energy contributions.
}
\end{figure}

The effective Hamiltonian $\ob{\Pminus}$ is constructed with
use of the Baker--Hausdorff expansion (\ref{eq:BHexpansion})
as applied to $\Pminus$ in (\ref{eq:Pminus}) and $T$ in (\ref{eq:T}).
The first commutator with the first term in $\Pminus$ is
\bea
[\Pminus_{0a},T]&=& \sum_{ijl\sigma s\lambda}
                \int dy d\veck \int \frac{d\ub{p}}{\sqrt{16\pi^3 p^+}}
          \left[\frac{\mu_{l\lambda}^2+(y\vec{p}_\perp+\veck)^2}{y}\right]
                \\
  &&  \times t_{ijl}^{\sigma a \lambda}(y,\veck) a_{l\lambda}^\dagger(y,\veck;\ub{p}) 
               b_{js}^\dagger(1-y,-\veck;\ub{p}) b_{i\sigma}(\ub{p}).
\eea
For the second term, we have
\bea
[\Pminus_{0b},T]&=& \sum_{ijl\sigma s\lambda}
                \int dy d\veck \int \frac{d\ub{p}}{\sqrt{16\pi^3 p^+}}
          \left[\frac{m_j^2+((1-y)\vec{p}_\perp-\veck)^2}{1-y}-(m_i^2+p_\perp^2)\right]
                \\
  &&  \times t_{ijl}^{\sigma a \lambda}(y,\veck) a_{l\lambda}^\dagger(y,\veck;\ub{p}) 
               b_{js}^\dagger(1-y,-\veck;\ub{p}) b_{i\sigma}(\ub{p}).
\nonumber
\eea
Due to cancellations between terms involving the transverse momentum $\vec{p}_\perp$,
the combination of the two simplifies to
\bea
[\Pminus_{0a}+\Pminus_{0b},T]&=& \sum_{ijl\sigma s\lambda}
                \int dy d\veck \int \frac{d\ub{p}}{\sqrt{16\pi^3 p^+}}
          \left[\frac{m_j^2+\veck^2}{1-y}+\frac{\mu_{l\lambda}^2+\veck^2}{y}-m_i^2\right]
                \\
  &&  \times t_{ijl}^{\sigma a \lambda}(y,\veck) a_{l\lambda}^\dagger(y,\veck;\ub{p}) 
               b_{js}^\dagger(1-y,-\veck;\ub{p}) b_{i\sigma}(\ub{p}).
\nonumber
\eea
A graphical representation is given in Fig.~\ref{fig:P0T}.
Since this commutator already involves a net increase by one particle, no additional
commutators with $T$ need to be considered for these terms, in the chosen
truncation.  Similarly, the photon emission term in $\Pminus_{\rm int}$
already increases the particle number and need not be considered at all
in any commutator.  For the photon absorption term, however, two commutators
must be considered.

\begin{figure}[ht]
\vspace{0.2in}
\centerline{\includegraphics[width=12cm]{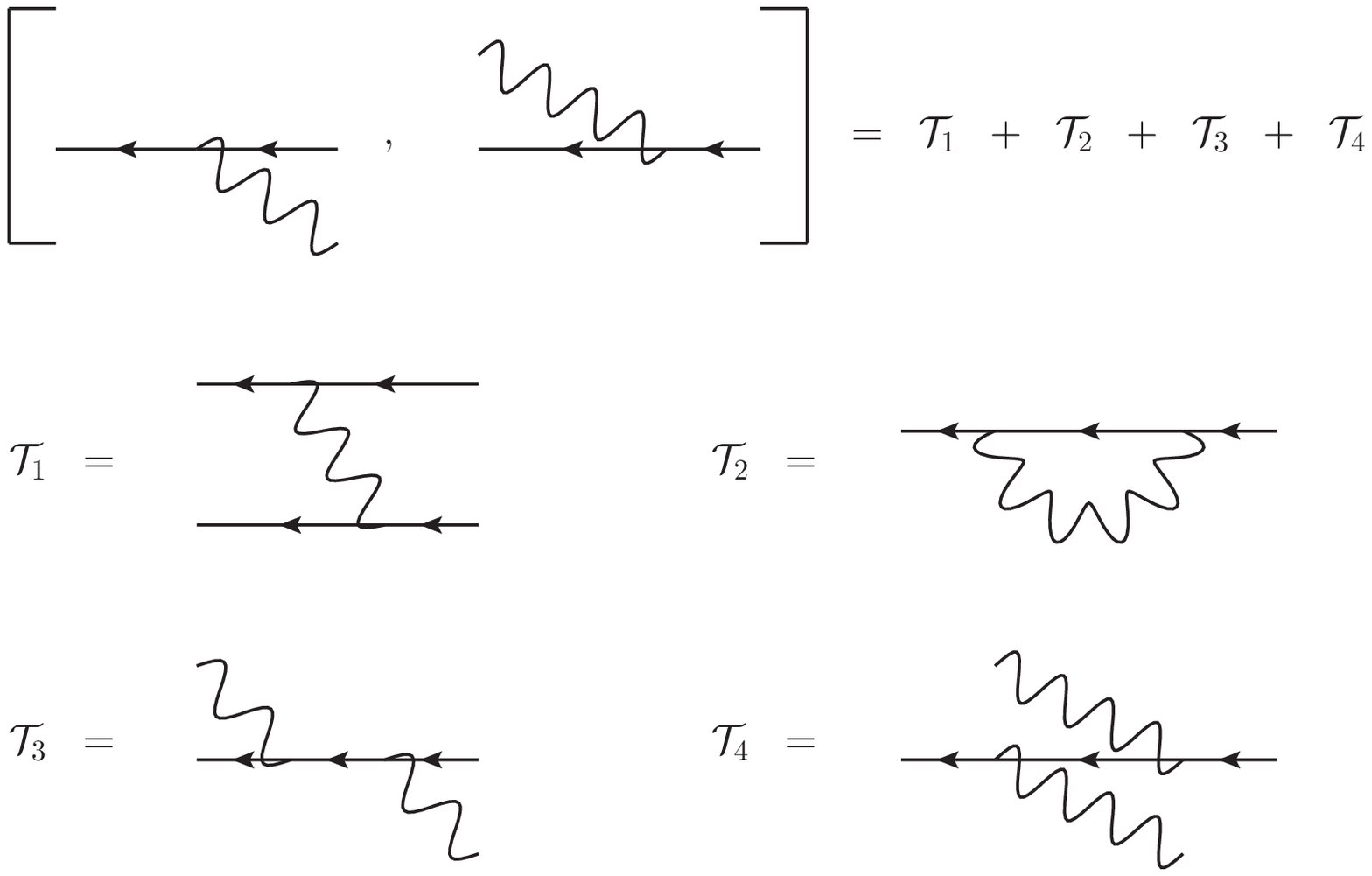}}
\caption{\label{fig:PintT} Graphical representation of
the terms in $[\Pminus_{\rm int},T]$.
}
\end{figure}

The first commutator with $\Pminus_{\rm int}$ generates terms that 
do not change particle number.  They can be represented graphically
as in Fig.~\ref{fig:PintT}.
The commutator can be written as
\be
[\Pminus_{\rm int},T]={\cal T}_1+{\cal T}_2+{\cal T}_3+{\cal T}_4,
\ee
with
\bea
{\cal T}_1&=& -\sum_{ijl\sigma s\lambda}(-1)^l\epsilon^\lambda
  \sum_{i'j'\sigma's'}\int dy d\veck \int \frac{d\ub{p}}{\sqrt{16\pi^3 p^+}}
  \int dy' d\veckp \int \frac{d\ub{p}'}{\sqrt{16\pi^3}}\sqrt{p^{\prime+}}  \\
  && \rule{0.5in}{0mm} \times h_{ijl}^{\sigma s \lambda *}(y,\veck)
                t_{i'j'l}^{\sigma' s'\lambda}(y',\veckp)
                \delta(yp^+-y'p^{\prime+})
                \delta(y\vec{p}_\perp+\veck-y'\vec{p}_\perp^{\,\prime}-\veckp) \nonumber \\
  && \rule{0.5in}{0mm}  \times b_{j's'}^\dagger(1-y',-\veckp;\ub{p}')b_{i\sigma}^\dagger(\ub{p})
                   b_{js}(1-y,-\veck;\ub{p}) b_{i'\sigma'}(\ub{p}'),
\nonumber \\
{\cal T}_2&=&\sum_{ij\sigma\sigma'}(-1)^i I_{ji}^{\sigma\sigma'}\int\frac{d\ub{p}}{p^+} b_{j\sigma}^\dagger(\ub{p})b_{i\sigma'}(\ub{p}), \\
{\cal T}_3&=& -\sum_{ijl\sigma s \lambda}(-1)^i \sum_{j'l's'\lambda'} 
                  \int dy d\veck \int dy' d\veckp \int \frac{d\ub{p}}{16\pi^3}
                  h_{ijl}^{\sigma s\lambda*}(y,\veck) t_{ij'l'}^{\sigma s'\lambda'}(y',\veckp)  \\
  && \rule{1in}{0mm} \times     a_{l'\lambda'}^\dagger(y',\veckp;\ub{p})
                  b_{j's'}^\dagger(1-y',-\veckp;\ub{p})
                  b_{js}(1-y,-\veck;\ub{p}) a_{l\lambda}(y,\veck;\ub{p}),
\nonumber \\
{\cal T}_4&=& \sum_{ijl\sigma s\lambda}(-1)^j 
             \int dy d\veck \int \frac{d\ub{p}}{\sqrt{16\pi^3 p^+}}
             \sum_{i'l'\sigma'\lambda'}
             \int dy' d\veckp \int \frac{d\ub{p}'}{\sqrt{16\pi^3}}\sqrt{p^{\prime+}}  \\
     && \rule{0.5in}{0mm} \times \delta((1-y)p^+-(1-y')p^{\prime+})
           \delta((1-y)\vec{p}_\perp-\veck-(1-y')\vec{p}'_\perp+\veckp) \nonumber \\
  && \rule{0.5in}{0mm}
   \times h_{ijl}^{\sigma s\lambda*}(y,\veck) t_{i'jl'}^{\sigma' s\lambda'}(y',\veckp)
           a_{l'\lambda'}^\dagger(y',\veckp;\ub{p}')b_{i\sigma}^\dagger(\ub{p})
               b_{i'\sigma'}(\ub{p}') a_{l\lambda}(y,\veck;\ub{p}),
\nonumber
\eea
and 
\be  \label{eq:Iji}
I_{ji}^{\sigma\sigma'}=(-1)^i\sum_{i'ls\lambda}(-1)^{i'+l}\epsilon^\lambda
        \int \frac{dy d\veck}{16\pi^3 }
        h_{ji'l}^{\sigma s\lambda*}(y,\veck) 
        t_{ii'l}^{\sigma' s\lambda}(y,\veck).
\ee
The first term, ${\cal T}_1$, does not contribute to the chosen truncation
to one fermion.
The spin dependence of the self-energy term ${\cal T}_2$ can be simplified
upon the observation that the dependence of $t_{ii'l}^{\sigma' s\lambda}(y,\veck)$
on the azimuthal angle of $\veck$ must follow the pattern of dependence 
in $h_{ii'l}^{\sigma' s\lambda}(y,\veck)$, in order that $T$ conserve $J_z$.
The azimuthal integral in $I_{ji}^{\sigma\sigma'}$ then implies that
\be
I_{ji}^{\sigma\sigma'}=\delta_{\sigma\sigma'}I_{ji},
\ee
with $I_{ji}$ real and independent of the spin projection $\sigma$.
This then simplifies ${\cal T}_2$ to be
\be
{\cal T}_2=\sum_{ijs}(-1)^i\int\frac{d\ub{p}}{p^+} I_{ji}b_{js}^\dagger(\ub{p})b_{is}(\ub{p}).
\ee

\begin{figure}[ht]
\vspace{0.2in}
\begin{tabular}{c}
\centerline{\includegraphics[width=15cm]{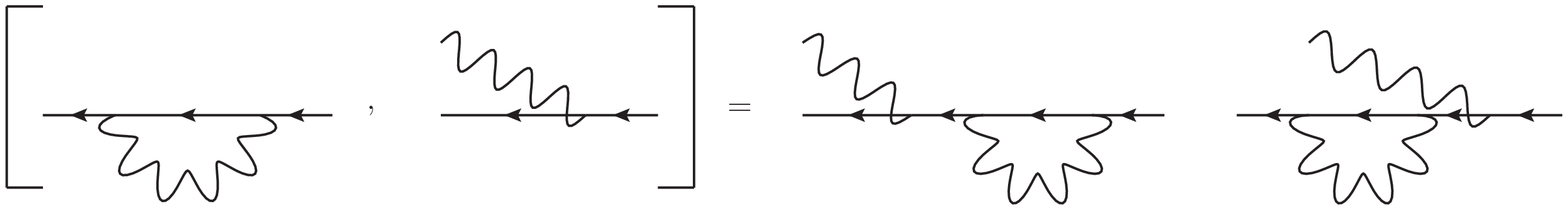}} \\
(a) \\
\centerline{\includegraphics[width=12cm]{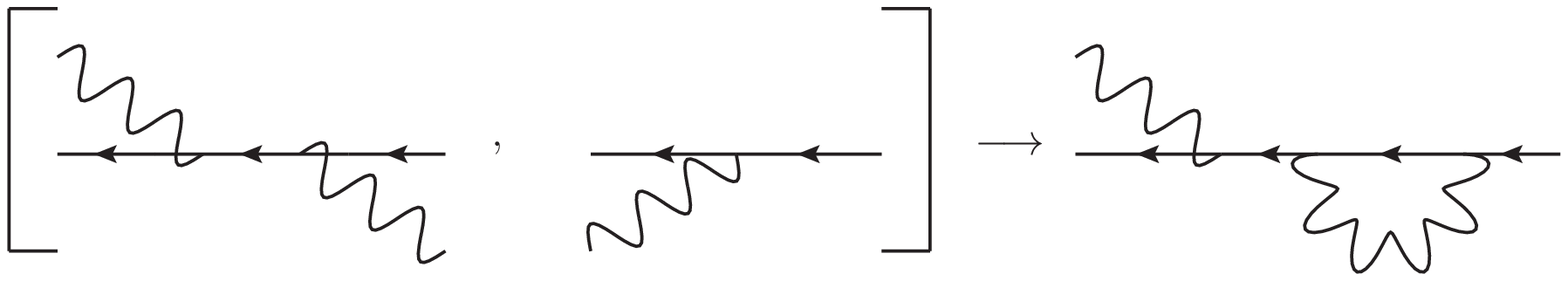}} \\
(b) \\
\centerline{\includegraphics[width=12cm]{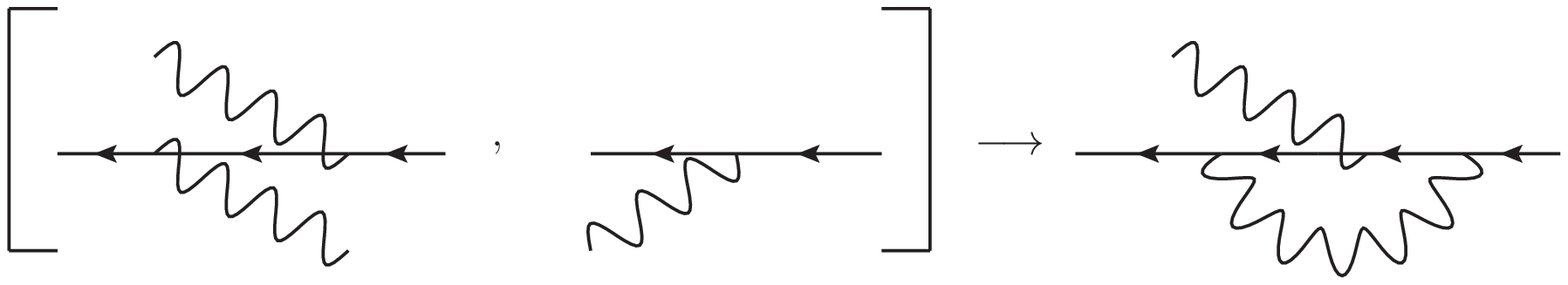}} \\
(c)
\end{tabular}
\caption{\label{fig:calTT} Graphical representation of
the contributions from $[[\Pminus_{\rm int},T],T]$,
with (a), (b), and (c) corresponding to $[{\cal T}_2,T]$,
$[{\cal T}_3,T]$, and $[{\cal T}_4,T]$, respectively.
$[{\cal T}_1,T]$ does not contribute at all for the
chosen truncation.}
\end{figure}

The second commutator $[[\Pminus_{\rm int},T],T]$ generates several
more terms, all of which increase particle number by one; however,
most of these do not contribute.  Graphical
representations are given in Fig.~\ref{fig:calTT}.
In particular, $[{\cal T}_1,T]$ 
contributes nothing for our truncation.  The next possibility
$[{\cal T}_2,T]$ does contribute fully and, of course, has a
structure very similar to that of $[\Pminus_{0b},T]$:
\bea
[{\cal T}_2,T]&=&\sum_{ijl\sigma s\lambda}
              \int dy d\veck \int \frac{d\ub{p}}{\sqrt{16\pi^3 p^+}}
              \left[ \sum_{i'}\frac{I_{ji'}}{1-y} t_{ii'l}^{\sigma s\lambda}(y,\veck)
                 -\sum_{j'}(-1)^{i+j'}t_{j'jl}^{\sigma s\lambda}(y,\veck)I_{j'i}\right] 
                 \nonumber \\
 &&  \rule{1in}{0mm} \times a_{l\lambda}^\dagger(y,\veck;\ub{p})
      b_{js}^\dagger(1-y,-\veck;\ub{p})b_{i\sigma}(\ub{p}).
\eea
The third term contains several pieces but contributes only
\bea
[{\cal T}_3,T]&\rightarrow& -\sum_{ijl\sigma s\lambda} 
\int dy d\veck \int \frac{d\ub{p}}{\sqrt{16\pi^3 p^+}} 
\sum_{j'} (-1)^{i+j'}t_{j'jl}^{\sigma s\lambda}(y,\veck) I_{j'i}\\
&&  \rule{1in}{0mm} \times a_{l\lambda}^\dagger(y,\veck;\ub{p})
      b_{js}^\dagger(1-y,-\veck;\ub{p})b_{i\sigma}(\ub{p}),
\nonumber
\eea
which combines with the second term of $[{\cal T}_2,T]$.
The fourth also contains several pieces; all that remains
in the given truncation is 
\be
[{\cal T}_4,T]\rightarrow \sum_{ijl\sigma s \lambda} 
\int dy d\veck \int \frac{d\ub{p}}{\sqrt{16\pi^3 p^+}} V_{ijl}^{\sigma s \lambda}(y,\veck)
a_{l\lambda}^\dagger(y,\veck;\ub{p})
      b_{js}^\dagger(1-y,-\veck;\ub{p})b_{i\sigma}(\ub{p}),
\ee%
where
\bea
V_{ijl}^{\sigma s\lambda}(y,\veck)
&=&\sum_{i'j'l'\sigma' s\lambda'}(-1)^{i'+j'+l'}\epsilon^{\lambda'}
  \int \frac{dy' d\veckp}{16\pi^3}
  \frac{\theta(1-y-y')}{(1-y')^{1/2}(1-y)^{3/2}} \\
&&\times 
  h_{jj'l'}^{ss'\lambda'*}(\frac{y'}{1-y},\veckp+\frac{y'}{1-y}\veck)
  t_{i'j'l}^{\sigma's'\lambda}(\frac{y}{1-y'},\veck+\frac{y}{1-y'}\veckp)
  t_{ii'l'}^{\sigma\sigma'\lambda'}(y',\veckp).  \nonumber
\eea
This is a vertex correction, as illustrated in Fig.~\ref{fig:calTT}(c).

\begin{figure}[ht]
\vspace{0.2in}
\centerline{\includegraphics[width=12cm]{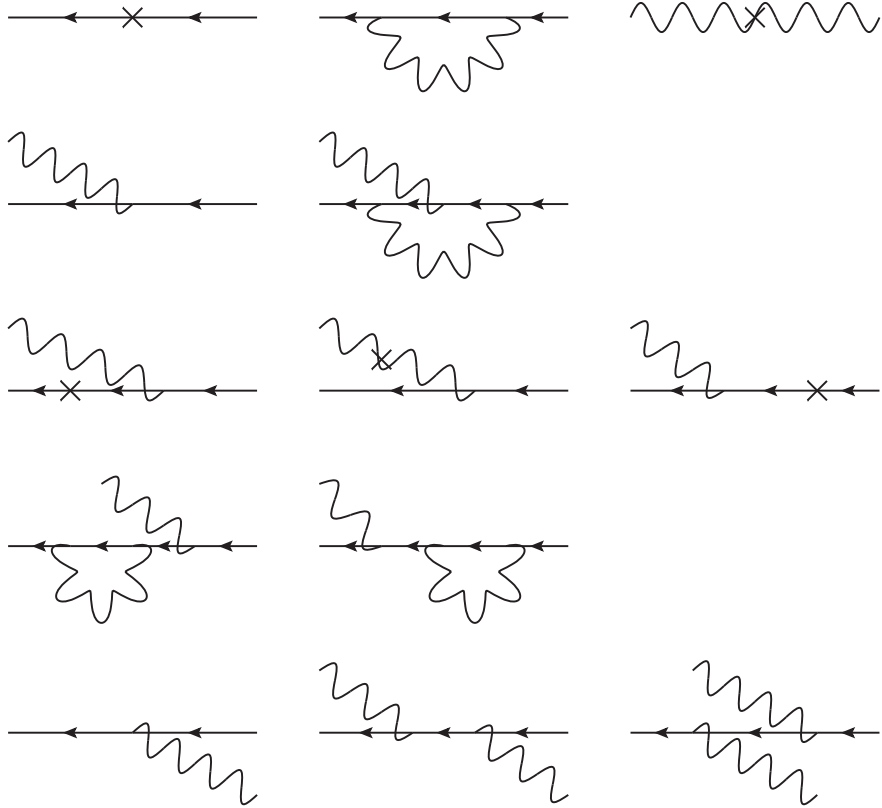}}
\caption{\label{fig:effH} Graphical representation
of the effective Hamiltonian $\ob{\Pminus}$, limited
to those terms that contribute in the given
truncation of the operator $T$.
}
\end{figure}

The effective Hamiltonian for the given truncation is then the 
sum of all these terms, according to the Baker--Hausdorff
expansion.  A graphical representation is shown in Fig.~\ref{fig:effH}.
The final expression is
\bea \label{eq:EffH}
\ob{\Pminus}&=&\sum_{ijs}(-1)^i\int d\ub{p}
      \left[\delta_{ij}\frac{m_i^2+p_\perp^2}{p^+}+\frac{I_{ji}}{p^+}\right]
          b_{js}^\dagger(\ub{p}) b_{is}(\ub{p}) \\
&& +\sum_{l\lambda}(-1)^l\epsilon^\lambda\int d\ub{p}
          \frac{\mu_{l\lambda}^2+p_\perp^2}{p^+}
             a_{l\lambda}^\dagger(\ub{p}) a_{l\lambda}(\ub{p}) \nonumber  \\
&&+\sum_{ijls\sigma\lambda}\int dy d\veck 
   \int\frac{d\ub{p}}{\sqrt{16\pi^3p^+}}
          \left\{\rule{0mm}{0.3in}h_{ijl}^{\sigma s\lambda}(y,\vec{k}_\perp)
            +\frac12 V_{ijl}^{\sigma s\lambda}(y,\vec k_\perp)\right. \nonumber \\
&&\rule{1.5in}{0mm}+\left[\frac{m_j^2+k_\perp^2}{1-y}
                                +\frac{\mu_{l\lambda}^2+k_\perp^2}{y}-m_i^2\right]
                                t_{ijl}^{\sigma s\lambda}(y,\vec k_\perp)\nonumber \\
&& \rule{1.5in}{0mm}\left.
    +\frac12\sum_{i'}\frac{I_{ji'}}{1-y}t_{ii'l}^{\sigma s\lambda}(y,\veck)
              -\sum_{j'}(-1)^{i+j'}t_{j'jl}^{\sigma s\lambda}(y,\veck)I_{j'i}\right\}
                                  \nonumber \\
&& \rule{1.5in}{0mm}\times  a_{l\lambda}^\dagger(y,\veck;\ub{p})
   b_{js}^\dagger(1-y,-\veck;\ub{p})b_{i\sigma}(\ub{p}) \nonumber \\
&&+\sum_{ijls\sigma\lambda}\int dy d\veck 
   \int\frac{d\ub{p}}{\sqrt{16\pi^3p^+}}
         h_{ijl}^{\sigma s\lambda*}(y,\vec{k}_\perp) b_{i\sigma}^\dagger(\ub{p})
   b_{js}(1-y,-\veck;\ub{p}) 
      a_{l\lambda}(y,\veck;\ub{p}) \nonumber \\
&& +\sum_{ijl\sigma s \lambda} (-1)^j\sum_{i'l'\sigma'\lambda'}
\int dy d\veck \int\frac{d\ub{p}}{\sqrt{16\pi^3 p^+}}
\int dy' d\veckp \int \frac{d\ub{p}'}{\sqrt{16\pi^3}}\sqrt{p^{\prime +}} \nonumber \\
&& \rule{0.4in}{0mm}\times 
\delta((1-y)p^+-(1-y')p^{\prime +})
\delta((1-y)\vec p_\perp-\veck-(1-y')\vec p_\perp^{\,\prime}+\veckp) \nonumber \\
&& \rule{0.4in}{0mm}\times 
h_{ijl}^{\sigma s\lambda*}(y,\veck)t_{i'jl'}^{\sigma' s\lambda'}(y',\veckp)
a_{l'\lambda'}^\dagger(y',\veckp;\ub{p'})
b_{i\sigma}^\dagger(\ub{p})b_{i'\sigma'}(\ub{p}')a_{l\lambda}(y,\veck;\ub{p})
    \nonumber \\
&& -\sum_{ijl\sigma s\lambda}(-1)^i \sum_{j'l's'\lambda'}
\int dy d\veck \int \frac{d\ub{p}}{16\pi^3} 
\int dy' d\veckp h_{ijl}^{\sigma s \lambda*}(y,\veck)
t_{ij'l'}^{\sigma s'\lambda'}(y',\veckp) \nonumber \\
&& \rule{0.4in}{0mm}\times a_{l'\lambda'}^\dagger(y',\veckp;\ub{p})
b_{j's'}^\dagger(1-y',-\veckp;\ub{p}) b_{js}(1-y,-\veck;\ub{p}) 
a_{l\lambda}(y,\veck;\ub{p}) \nonumber
\eea
%


\end{document}